\documentclass[aps,prl,twocolumn,superscriptaddress,nofootinbib,longbibliography]{revtex4-2}

\usepackage{graphicx}
\usepackage{dcolumn}
\usepackage{amsmath}
\usepackage{amssymb}
\usepackage{booktabs}
\usepackage{xcolor}
\usepackage{multirow}
\usepackage{bm}
\usepackage{mhchem}
\usepackage[hidelinks]{hyperref}
\usepackage{float}

\begin{document}

\title{Interplay between Spin Stripes and Pairing Crossover in the Bilayer Hubbard Model}

\author{Zixuan Jia}
\affiliation{School of Physics and Astronomy, and Key Laboratory of Multiscale Spin Physics (Ministry of Education), Beijing Normal University, Beijing 100875, China}
\affiliation{Beijing Computational Science Research Center, Beijing 100193, China}
\author{Xuelei Sui}
\affiliation{Beijing Computational Science Research Center, Beijing 100193, China}
\affiliation{College of Mathematics and Physics, Beijing University of Chemical Technology, Beijing 100029, China}
\author{Shijie Hu}
\email{shijiehu@csrc.ac.cn}
\affiliation{School of Physics and Astronomy, and Key Laboratory of Multiscale Spin Physics (Ministry of Education), Beijing Normal University, Beijing 100875, China}
\affiliation{Beijing Computational Science Research Center, Beijing 100193, China}
\author{Tianxing Ma}
\email{txma@bnu.edu.cn}
\affiliation{School of Physics and Astronomy, and Key Laboratory of Multiscale Spin Physics (Ministry of Education), Beijing Normal University, Beijing 100875, China}
\author{Haiqing Lin}
\affiliation{School of Physics and Institute for Advanced Studies of Physics, Zhejiang University, Hangzhou, 310058, China}
\author{Bing Huang}
\email{bing.huang@csrc.ac.cn}
\affiliation{School of Physics and Astronomy, and Key Laboratory of Multiscale Spin Physics (Ministry of Education), Beijing Normal University, Beijing 100875, China}
\affiliation{Beijing Computational Science Research Center, Beijing 100193, China}

\begin{abstract}
Whether $d$-wave or $s_\pm$-wave pairing prevails in the bilayer nickelate \ce{La3Ni2O7} remains unresolved, and how its observed spin order and magnetic excitations relate to the pairing mechanism is unclear. 
Here, we use finite-temperature determinant quantum Monte Carlo simulations of a minimal bilayer Hubbard model to investigate pairing competition at strong interlayer coupling, where intralayer and interlayer antiferromagnetic correlations have comparable strengths.
We find that spin stripes develop spontaneously with increasing hole doping, accompanied by a crossover of the dominant pairing symmetry from $d$-wave to $s_\pm$-wave.
The crossover tracks the changing relative strengths of the spin correlations, with the interlayer correlations overtaking their intralayer counterparts in magnitude.
Concomitantly, the dynamical spin structure factor exhibits a momentum-dependent reorganization of the magnetic excitations, consistent with fluctuating spin stripes.
Our results identify the structure of spin fluctuations as a key link between spin stripes and pairing competition, offering a minimal-model perspective for understanding \ce{La3Ni2O7}.
\end{abstract}

\maketitle

\paragraph{Introduction---}\hspace{-1em}
The discovery of superconductivity in pressurized \ce{La3Ni2O7}~\cite{327SCSun2023,327SCYuan2024,327SCWang2024} has opened a new frontier in the search for high-$T_c$ superconductivity beyond the cuprates. 
Determining the symmetry of the superconducting gap is central to identifying the pairing mechanism, yet experiments remain divided between $d$-wave and $s_\pm$-wave scenarios~\cite{327PCSCao2025,327PCSLiu2025,327PCSGuo2025}. 
Recent observations of nodeless superconducting gaps in ambient-pressure bilayer nickelate thin films lend further support to the $s_{\pm}$-wave scenario~\cite{327thinfilmShen2026,327thinfilmWen2026}, deepening the ongoing debate. 
A recurring theme in unconventional superconductors is the role of spin fluctuations in the pairing mechanism.
In hole-doped cuprates, antiferromagnetic spin fluctuations are extensively documented~\cite{CuflucTranquada1995,CuflucHayden2004,CuflucYamada2004,CuflucHayden2007,CuAFMTranquada2007} and widely regarded as central to the well-established $d$-wave pairing~\cite{CutheoScalapino1986,CutheoMonthoux1991,CutheoScalapino1995,CutheoMoriya2000,CutheoAbanov2008,pairreviewScalapino2012}. 
In bilayer systems, an ultracold-atom experiment has shown that increasing interlayer coupling suppresses intralayer antiferromagnetic correlations while promoting interlayer spin-singlet formation~\cite{bicoldGali2021}, and theoretical studies associate the competition between these two fluctuation channels with the selection between $d$-wave and $s_\pm$-wave pairing~\cite{biRichard1994,bipairSantos1995,bipairAndersen1995,bipairLee2009,bidspairScalapino2011}.
Given the substantial interlayer coupling identified in \ce{La3Ni2O7}~\cite{327interWang2024,327SDWRIXSChen2024,327spinfulcChen2026}, this competition may well reshape the magnetic response and tip the balance between the two pairing symmetries.

Experiments have revealed spin-density-wave (SDW) order in \ce{La3Ni2O7}~\cite{327SDWRIXSChen2024,327SDWuSRChen2024,327SDWNMRZhao2025,327SDWuSRKhasanov2024,327spinfulcChen2026}, with a stripe-type spin structure proposed as a likely configuration.
Numerical simulations of single-layer Hubbard models have characterized stripe configurations in the ground state, including their coexistence with $d$-wave superconductivity~\cite{stripeGSZheng2017,stripeGSttXu2024,stripedwaveHu2026,stripeYang2026}.
Finite-temperature studies further uncover fluctuating stripes with characteristic magnetic excitations, which may alter the dominant pairing symmetry~\cite{stripeDQMCEdwin2018,stripeDQMC3bandEdwin2017,stripeDQMCEdwin2023,stripepairChen2024}.
However, how stripes interact with superconducting pairing at finite temperature remains essentially unexplored in bilayer systems where intralayer and interlayer spin fluctuations compete on comparable footing.

Various models have been proposed for the electronic structure and superconductivity of \ce{La3Ni2O7}, notably bilayer multiorbital models that incorporate the Ni $3d_{x^2-y^2}$ and $3d_{z^2}$ orbitals and their interactions~\cite{327bibiYao2023,327bibiWang2023,327bibiChen2025,327pairXiong2026}.
Yet the essential electronic ingredients for superconductivity remain under debate: recent ARPES studies of superconducting bilayer nickelate films suggest that the $d_{z^2}$-derived $\gamma$ pocket may not be a prerequisite~\cite{327gammaWang2026,327gammaSun2026}.
Simplified bilayer models have meanwhile offered complementary insights into how interlayer coupling and electronic correlations jointly shape the superconducting pairing~\cite{327biXiang2024,327bitJJQu2024,327biZhang2024}. 
Among them, the single-orbital bilayer Hubbard model retains two essential ingredients---local Coulomb interaction and interlayer hybridization~\cite{bidspairScalapino2011}---and thus provides a minimal framework for studying spin fluctuations and pairing competition in bilayer systems.

\begin{figure}[tbp]
\includegraphics[width=\columnwidth]{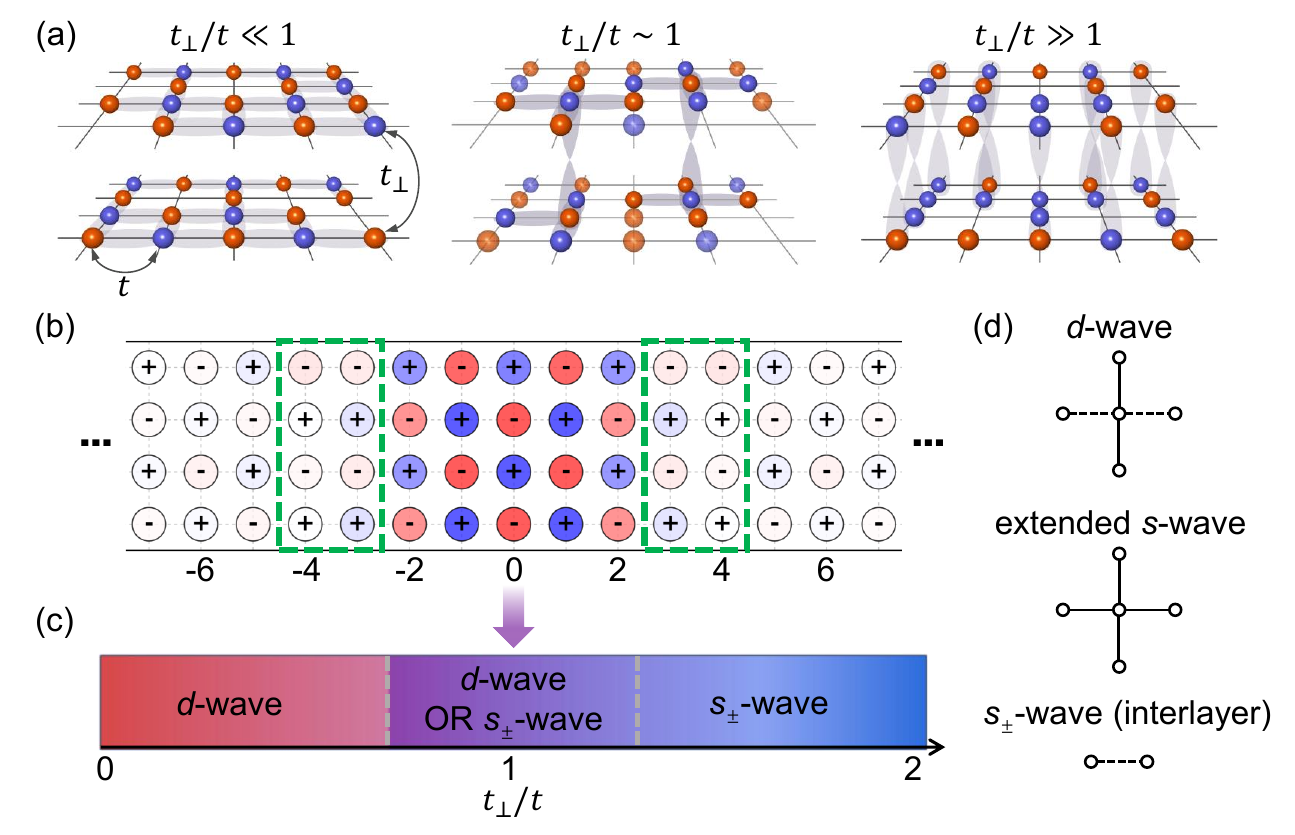}
\caption{
(a) Schematic spin correlations in the bilayer Hubbard model at different interlayer hopping strengths. Red (blue) spheres denote spin-up (spin-down) electrons. Intralayer antiferromagnetic correlations dominate at $t_\perp/t\ll1$, whereas interlayer antiferromagnetic correlations dominate at $t_\perp/t\gg1$; near $t_\perp/t\sim1$, the two compete with comparable strengths. Only representative correlations are shown for clarity.
(b) Spin stripes in the real-space spin susceptibility upon hole doping. Blue (red) circles marked with $+$ ($-$) denote positive (negative) values, and green dashed lines mark the antiphase domain walls.
(c) Schematic evolution of the dominant pairing symmetry with $t_\perp/t$. Near $t_\perp/t\sim1$, spin stripes develop upon hole doping, associated with a crossover from $d$-wave to $s_\pm$-wave pairing. The vertical dashed lines are schematic boundaries.
(d) Real-space form factors of the $d$-wave, extended $s$-wave, and interlayer $s_\pm$-wave pairing symmetries. Dashed (solid) bonds denote positive (negative) form factors.
}
\label{fig1}
\end{figure}

In this Letter, we employ determinant quantum Monte Carlo (DQMC) simulations to investigate spin correlations and superconducting pairing in the single-orbital bilayer Hubbard model at finite temperature. 
We focus on the regime $t_\perp/t\sim1$, where intralayer and interlayer antiferromagnetic correlations have comparable strengths, as illustrated schematically in Fig.~\ref{fig1}(a).
Upon hole doping, spin stripes develop spontaneously [Fig.~\ref{fig1}(b)], accompanied by a crossover of the dominant pairing symmetry from $d$-wave to $s_{\pm}$-wave [Fig.~\ref{fig1}(c)].
The crossover tracks the changing relative strengths of the two spin-correlation channels, with the interlayer correlations overtaking their intralayer counterparts in magnitude.
The dynamical spin structure factor exhibits a concomitant momentum-dependent reorganization of the magnetic excitations, consistent with fluctuating spin stripes.
Our results reveal the interplay between spin stripes and pairing competition in bilayer systems, identifying the importance of the structure of spin fluctuations. 

\paragraph{Model and method---}\hspace{-1em}
We consider the single-orbital bilayer Hubbard model described by the Hamiltonian
\begin{equation}
\begin{aligned}
\hat{H} ={} 
& -t\sum_{\langle ij\rangle \ell\sigma}
\hat{c}_{i\ell\sigma}^{\dagger}
\hat{c}_{j\ell\sigma}
-t'\sum_{\langle\langle ij\rangle\rangle \ell\sigma}
\hat{c}_{i\ell\sigma}^{\dagger}
\hat{c}_{j\ell\sigma}
\\
&-t_{\perp}\sum_{i\sigma}
\hat{c}_{i1\sigma}^{\dagger}
\hat{c}_{i2\sigma}
+U\sum_{i\ell}
\hat{n}_{i\ell\uparrow}
\hat{n}_{i\ell\downarrow} 
-\mu\sum_{i\ell\sigma}
\hat{n}_{i\ell\sigma} ,
\end{aligned}
\end{equation}
where $\hat{c}_{i\ell\sigma}^{\dagger}$ ($\hat{c}_{i\ell\sigma}$) creates (annihilates) an electron with spin $\sigma=\uparrow,\downarrow$ at site $i$ in layer $\ell=1,2$, and $\hat{n}_{i\ell\sigma}=\hat{c}_{i\ell\sigma}^{\dagger}\hat{c}_{i\ell\sigma}^{}$ is the particle-number operator.
The hopping amplitudes $t$, $t^\prime$, and $t_\perp$ connect intralayer nearest-neighbor sites $\left\langle ij\right\rangle$, intralayer next-nearest-neighbor sites $\langle\langle ij\rangle\rangle$, and vertically aligned interlayer sites, respectively.
$U$ is the on-site Coulomb interaction and $\mu$ is the chemical potential.

We compute the pairing susceptibility in each symmetry,
\begin{equation}
P_{\alpha} =
\int_{0}^{\beta}d\tau
\left\langle
\hat{\Delta}_{\alpha}^{\dagger}(\tau)
\hat{\Delta}_{\alpha}(0)
\right\rangle ,
\end{equation}
where $\hat{\Delta}_{\alpha}^{\dagger}(\tau) = e^{\tau \hat{H}}\hat{\Delta}_{\alpha}^{\dagger}e^{-\tau \hat{H}}$. 
We consider three pairing symmetries: $d$-wave, extended $s$-wave (ES), and interlayer $s_{\pm}$-wave.
The corresponding order parameters read
\begin{equation}
\begin{gathered}
\hat{\Delta}_{\alpha}^{\dagger}=
\sum_{i\ell}
\sum_{\boldsymbol{\delta}}
f_{\alpha}(\boldsymbol{\delta})
\left(
\hat{c}_{i\ell\uparrow}^{\dagger}
\hat{c}_{i+\boldsymbol{\delta},\ell\downarrow}^{\dagger}
-
\hat{c}_{i\ell\downarrow}^{\dagger}
\hat{c}_{i+\boldsymbol{\delta},\ell\uparrow}^{\dagger}
\right),
\alpha=d,\mathrm{ES},
\\
\hat{\Delta}_{s_{\pm}}^{\dagger}=
\sum_i
\left(
\hat{c}_{i1\uparrow}^{\dagger}
\hat{c}_{i2\downarrow}^{\dagger}
-
\hat{c}_{i1\downarrow}^{\dagger}
\hat{c}_{i2\uparrow}^{\dagger}
\right).
\end{gathered}
\end{equation}

Here, $\boldsymbol{\delta}=\pm\hat{\mathbf{x}},\pm\hat{\mathbf{y}}$ denotes the in-plane nearest-neighbor vectors, and the real-space form factors of the three symmetries are illustrated in Fig.~\ref{fig1}(d).
To assess the interaction contribution to the pairing, we also compute the uncorrelated pair-field susceptibility $\overline{P}_{\alpha}$, obtained by factorizing the two-particle correlation function into products of single-particle Green's functions, and define the pairing vertex $\Gamma_{\alpha}=1/P_{\alpha}-1/\overline{P}_{\alpha}$~\cite{DQMCpairRubem2012}.
The effective pairing interaction in symmetry $\alpha$ is characterized by the dimensionless quantity $\Gamma_\alpha\overline{P}_\alpha$: a negative value indicates an attractive effective interaction, while $\Gamma_{\alpha}\overline{P}_{\alpha}\rightarrow -1$ signals an approaching pairing instability.

\begin{figure*}[t]
\includegraphics[width=\textwidth]{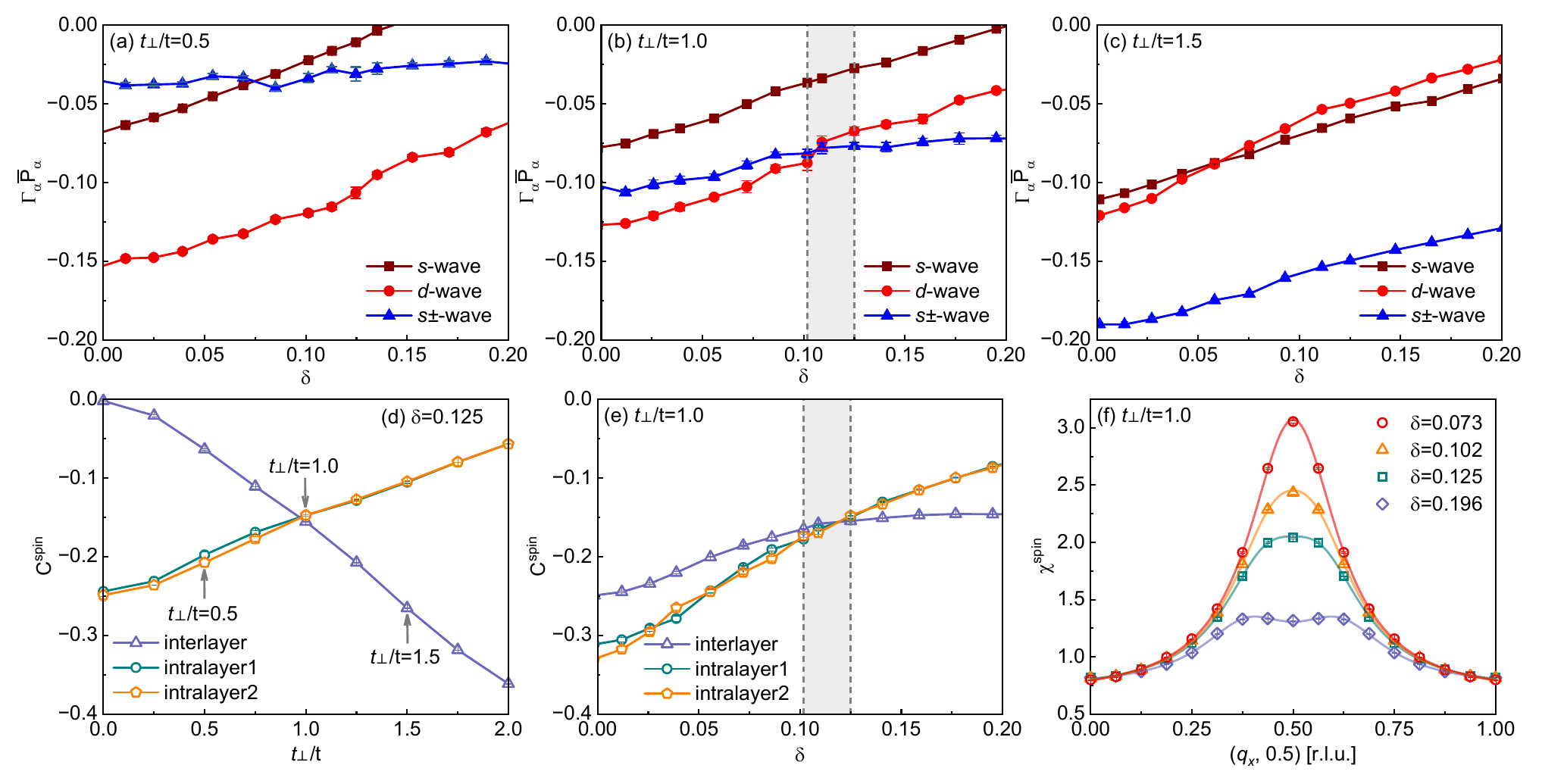}
\caption{
(a)--(c) Effective pairing interaction $\Gamma_\alpha\overline P_\alpha$ of the $d$-wave, extended $s$-wave, and $s_\pm$-wave versus hole doping $\delta$ at $t_\perp/t=0.5$, $1.0$, and $1.5$, respectively. Negative values indicate attractive effective pairing interactions. 
(d) Intralayer and interlayer spin correlations $C^{\mathrm{spin}}$ versus $t_\perp/t$ at $\delta=0.125$, with the intralayer correlations shown separately for the two layers.
(e) Doping dependence of the spin correlations at $t_\perp/t=1.0$. 
The shaded interval $0.102\lesssim\delta\lesssim0.125$ in (b) and (e) marks the doping range over which the pairing and spin-correlation strengths are compared near their respective crossovers.
(f) Intralayer spin susceptibility $\chi_{11}^{\mathrm{spin}}(\mathbf q)$ along $\mathbf q=(q_x,0.5)$ in reciprocal lattice units at $t_\perp/t=1.0$ for the indicated dopings. 
The two layers give nearly identical results, so only layer 1 is shown.
Symbols denote DQMC data, and lines are fits to single or double Lorentzian peaks.
Error bars are omitted when they are smaller than the symbols.
}
\label{fig2}
\end{figure*}

We calculate the real-space spin susceptibility to characterize spin stripes,
\begin{equation}
\chi^{\mathrm{spin}}_{\ell\ell'}(\mathbf r)=
\frac{1}{N_{xy}}
\sum_i\int_0^\beta d\tau\,
\left\langle
\hat{m}^z_{i+\mathbf r,\ell}(\tau)
\hat{m}^z_{i\ell'}(0)
\right\rangle ,
\end{equation}
where $\hat{m}^z_{i\ell}=\frac{1}{2}\left(\hat{n}_{i\ell\uparrow}-\hat{n}_{i\ell\downarrow}\right)$ is the local spin density, $\mathbf r$ is the spatial displacement, and $N_{xy}$ is the number of sites in each layer.
Its Fourier transform $\chi^{\mathrm{spin}}_{\ell\ell'}(\mathbf q)$ resolves the spin response in momentum space.
The nearest-neighbor intralayer and interlayer spin correlations are defined as
\begin{equation}
\begin{gathered}
C_{\mathrm{intra},\ell}^{}=
\frac{1}{4}
\sum_{\boldsymbol{\delta}=\pm\hat{\mathbf{x}},\pm\hat{\mathbf{y}}}
\chi _{\ell \ell}^{\mathrm{spin}}(\boldsymbol{\delta}),
\qquad
C_{\mathrm{inter}}
=
\chi_{12}^{\mathrm{spin}}(\boldsymbol{0}).
\end{gathered}
\end{equation}

All observables are evaluated using DQMC simulations~\cite{DQMCSantos2003}.
At finite temperature, sufficiently large clusters are required to accommodate multiple antiphase domain walls and capture spin stripes. 
Clear stripe signatures have been established in DQMC studies of the single-layer Hubbard model on $16\times4$ clusters, corroborated by finite-size comparisons and density matrix renormalization group calculations~\cite{stripeDQMCEdwin2018,stripeDQMC3bandEdwin2017,stripeDQMCEdwin2023}. 
Balancing the computational cost against the fermion-sign problem at larger sizes, we use clusters of $N=16\times4\times2$ sites with periodic boundary conditions within each layer. 
The intralayer hopping $t$ sets the energy unit.
Unless otherwise noted, the results are obtained for $U/t=4$, $t'/t=-0.25$, and $T/t=0.2$, at several values of $t_{\perp}/t$ and hole dopings $0\leq\delta\leq0.2$. 
The imaginary-time step $\Delta\tau$ is chosen such that the Trotter errors are below the statistical uncertainties. 
For each parameter set, we run 20--50 independently seeded Markov chains, each equilibrated for $4\times10^4$ sweeps followed by $5\times10^4$ measurement sweeps grouped into 5--10 bins. 
For all parameter sets shown in the main text, the average fermion sign satisfies $\langle\mathrm{sign}\rangle>0.1$.
Finite-size analyses and additional results for other interaction strengths and temperatures are presented in the End Matter.

\paragraph{Results and discussion---}\hspace{-1em}
To identify the regime where $d$-wave and $s_\pm$-wave pairing compete, we first compare the effective pairing interactions $\Gamma_\alpha\overline{P}_\alpha$ at three interlayer hopping strengths, $t_\perp/t=0.5$, $1.0$, and $1.5$ [Figs.~\ref{fig2}(a)--\ref{fig2}(c)].
At $t_\perp/t=0.5$, the $d$-wave pairing is the most attractive throughout the investigated doping range.
Upon increasing the interlayer hopping to $t_\perp/t=1.5$, the dominant pairing switches from $d$-wave to $s_\pm$-wave.
Figure~\ref{fig2}(d) reveals that this switch is accompanied by a reversal of the relative strengths of the intralayer and interlayer antiferromagnetic correlations, shown here at the representative doping $\delta=0.125$.
The evolution of the dominant pairing with increasing interlayer hopping is qualitatively consistent with previous studies of the bilayer Hubbard model~\cite{bidspairScalapino2011}.

We now focus on $t_\perp/t=1.0$, where the intralayer and interlayer antiferromagnetic correlations have comparable strengths and the dominant pairing symmetry changes with hole doping.
At low hole doping, the $d$-wave effective pairing interaction is the most attractive [Fig.~\ref{fig2}(b)].
With increasing doping, the $s_\pm$-wave pairing takes over as the dominant pairing.
The extended $s$-wave pairing, included for completeness, remains subleading throughout the parameter range.
This crossover tracks the changing relative strengths of the spin correlations, with the interlayer correlations overtaking their intralayer counterparts in magnitude [Fig.~\ref{fig2}(e)].
Both magnetic and pairing crossovers occur within the shaded doping interval $0.102\lesssim\delta\lesssim0.125$, where the spin stripes also emerge as we will discuss later.

As shown in Fig.~\ref{fig2}, $\Gamma_\alpha\overline{P}_\alpha<0$ for all pairing symmetries considered, indicating an effective attractive interaction, and its magnitude $|\Gamma_\alpha\overline{P}_\alpha|$ decreases monotonically with hole doping. 
This trend is qualitatively consistent with earlier studies of the repulsive Hubbard model, which attribute the pairing interaction largely to antiferromagnetic spin-fluctuation exchange~\cite{pairreviewScalapino2012}. 
This exchange is strongest near half filling, where the Fermi surface is approximately nested at $\mathbf{q}=(\pi,\pi)$, and weakens with hole doping. 
The decrease of $|\Gamma_\alpha\overline{P}_\alpha|$ may therefore reflect the weakening of this exchange. 
The suppression of antiferromagnetic correlations with hole doping [Fig.~\ref{fig2}(e)] parallels the reduction of the effective pairing strength, highlighting the close link between magnetic fluctuations and pairing correlations. 
Moreover, the interplay between spin-stripe order and electron interactions leads to a more complex magnetic reorganization and a pairing crossover upon hole doping as we discuss in this work.
We use $\Gamma_\alpha\overline{P}_\alpha$ to characterize the effective pairing strength in each symmetry $\alpha$. 
The superconducting transition temperature $T_c$, however, may also be affected by the pseudogap and the phase stiffness~\cite{pairTcLee2006}, and needs further exploration.

\begin{figure}[tbp]
\includegraphics[width=\columnwidth]{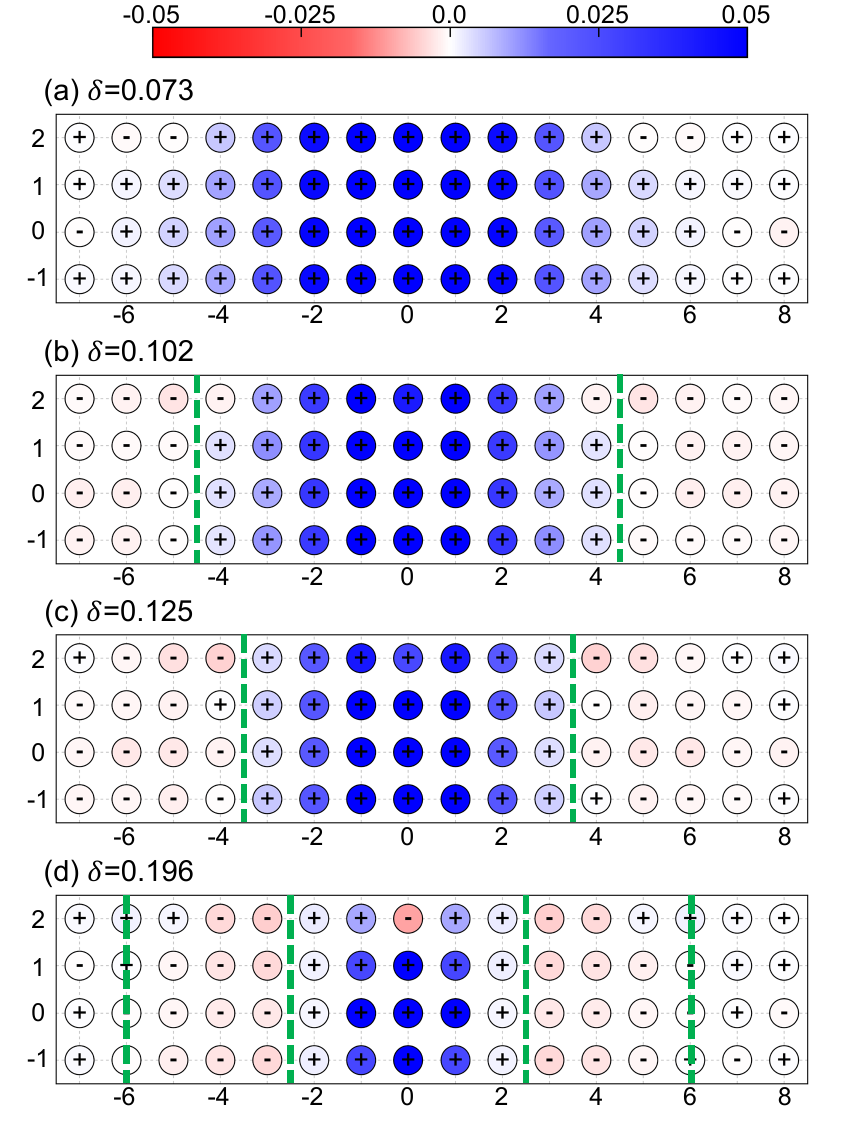}
\caption{
Staggered real-space spin susceptibility $\overline{\chi} ^{\mathrm{spin}}_{11}(\mathbf r)$ at hole dopings (a) $\delta=0.073$, (b) $0.102$, (c) $0.125$, and (d) $0.196$ for $t_\perp/t=1.0$. 
The two layers give nearly identical results, so only layer 1 is shown.
Blue (red) circles marked with $+$ ($-$) denote positive (negative) values.
All panels share the same color scale, and green dashed lines mark the antiphase domain walls. 
}
\label{fig3}
\end{figure}

We first examine the momentum-space spin susceptibility shown in Fig.~\ref{fig2}(f).
At $\delta=0.073$, the system exhibits antiferromagnetic correlations, with the spin susceptibility reaching its maximum at $(\pi,\pi)$. 
Within the crossover interval $0.102\lesssim\delta\lesssim0.125$, the spin susceptibility peak around $(\pi,\pi)$ broadens without developing a clearly resolved double-peak structure.
At the higher doping $\delta=0.196$, a shallow double-peak structure becomes visible.
Nevertheless, signatures of spin stripes are already visible within the crossover interval in the real-space spin susceptibility.
We multiply $\chi^{\mathrm{spin}}_{11}(\mathbf r)$ by the $(\pi,\pi)$ phase factor $(-1)^{r_x+r_y}$, yielding the staggered spin susceptibility shown in Fig. 3.
At $\delta=0.073$, the staggered susceptibility is predominantly positive without clear antiphase domains [Fig.~\ref{fig3}(a)], consistent with commensurate antiferromagnetic correlations.
At $\delta=0.102$, however, antiphase domains already appear as spatially organized sign reversals [Fig.~\ref{fig3}(b)], providing clear real-space evidence for the emergence of spin stripes in the crossover regime. 
With increasing hole doping, the spacing between antiphase domain walls shrinks, and multiple sets of domain walls coexist within the cluster at $\delta=0.196$ [Figs.~\ref{fig3}(c) and \ref{fig3}(d)].
The absence of resolved splitting is compatible with these real-space signatures, since a small incommensurate peak separation can be masked by the finite peak width.
This doping evolution of the spin stripes echoes previous finite-temperature findings for the single-layer Hubbard model~\cite{stripeDQMCEdwin2018,stripeDQMC3bandEdwin2017}.
Importantly, the doping range over which the spin stripes emerge coincides with the crossover regime identified in Fig.~\ref{fig2}, where both the dominant pairing symmetry and the structure of spin fluctuations undergo a crossover. 
This correspondence suggests that the emergence of spin stripes with hole doping may be accompanied by a reorganization of the spin-fluctuation structure, which in turn may be closely related to the crossover of the dominant pairing symmetry.

\begin{figure}[tbp]
\includegraphics[width=\columnwidth]{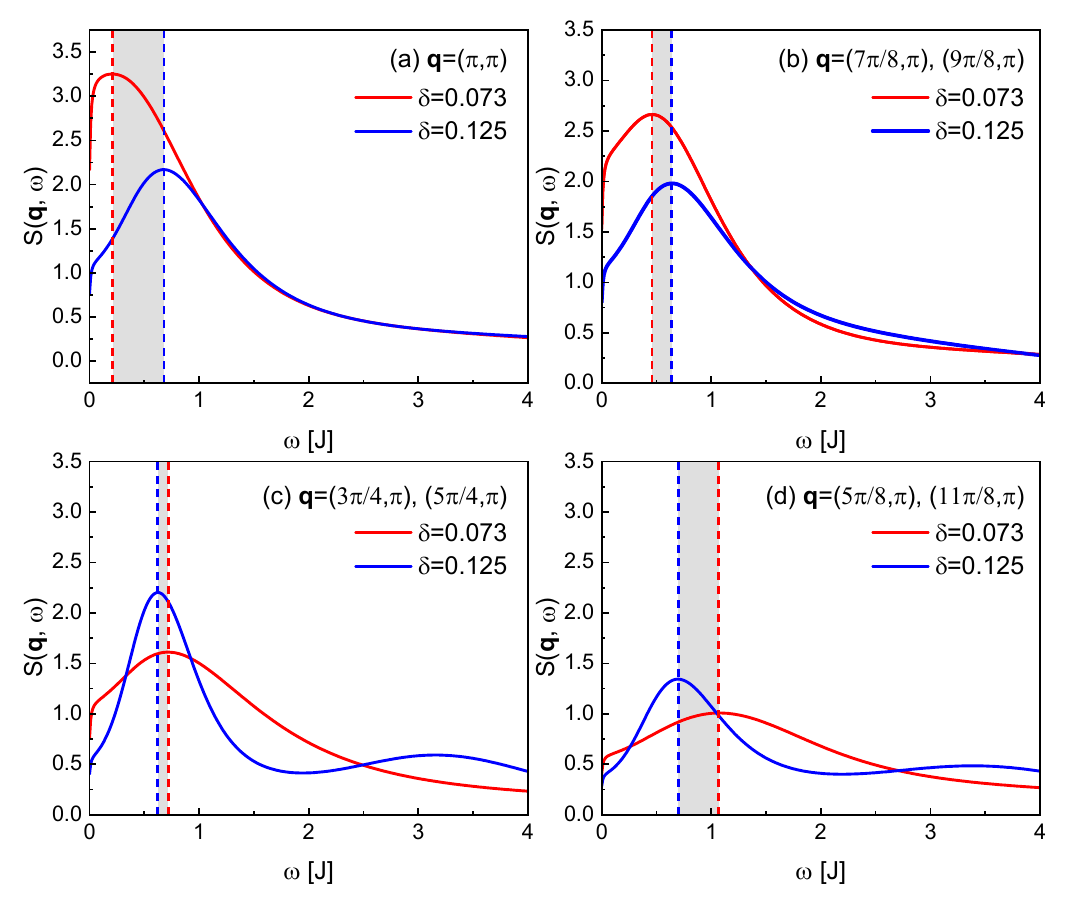}
\caption{
Odd channel of the dynamical spin structure factor, $S(\mathbf q,\omega)$, along $q_y=\pi$ at $t_\perp/t=1.0$ for $\delta=0.073$ and $0.125$. 
The spectra are obtained by maximum-entropy analytic continuation of the DQMC imaginary-time spin susceptibility. 
Vertical dashed lines mark the principal peak energies, and frequencies are in units of $J=4t^2/U$. 
}
\label{fig4}
\end{figure}

We next examine the magnetic excitations associated with the spin stripes.
Given the antiferromagnetic interlayer correlations, we analyze the odd channel of the dynamical spin structure factor, $S(\mathbf q,\omega)=S_{11}+S_{22}-S_{12}-S_{21}$, where $S_{\ell\ell'}$ denotes the layer-resolved spin structure factor and $\mathbf q$ is the in-plane momentum.
The spectra are extracted from the DQMC imaginary-time spin susceptibility via analytic continuation using the maximum entropy method~\cite{MEMJarrell1996,MEMMacridin2004}. 
Figure~\ref{fig4} compares $S(\mathbf q,\omega)$ along $q_y=\pi$ at $\delta=0.073$ and $0.125$ for $t_\perp/t=1.0$.
At $\mathbf q=(\pi,\pi)$, hole doping shifts the main spectral peak to higher energy and reduces its intensity.
The peaks at the neighboring momenta $(7\pi/8,\pi)$ and $(9\pi/8,\pi)$ also harden and lose intensity, though with smaller energy shifts.
By contrast, the peaks at $(3\pi/4,\pi)$ and $(5\pi/4,\pi)$ soften slightly, narrow, and gain intensity, and the softening becomes more pronounced at $(5\pi/8,\pi)$ and $(11\pi/8,\pi)$.
These changes in peak energy and intensity constitute a momentum-dependent reorganization of the magnetic excitations with hole doping.
Together with the antiphase domains in Fig.~\ref{fig3}, the spectral evolution is qualitatively consistent with the picture of the fluctuating spin stripes established in finite-temperature studies of single-layer Hubbard models~\cite{stripeDQMC3bandEdwin2017,stripeDQMCEdwin2018,stripeDQMCEdwin2023}.

In cuprates, static stripe order has been associated with a reduced $T_c$, while stripe-related magnetic responses persist outside the ordered phase~\cite{CuAFMTranquada2007}, motivating broad interest in fluctuating stripes and their relationship with superconductivity.
Finite-temperature studies of Hubbard models have shown that antiphase correlations and stripe-related magnetic excitations indeed persist in a fluctuating-stripe state~\cite{stripeDQMC3bandEdwin2017,stripeDQMCEdwin2018}.
In our bilayer model, the real-space antiphase correlations and the accompanying spectral reorganization lend qualitative support to the same picture.

\paragraph{Summary---}\hspace{-1em}
Using finite-temperature DQMC simulations of the bilayer Hubbard model at strong interlayer coupling, we find that the dominant pairing symmetry crosses over from $d$-wave to $s_\pm$-wave as spin stripes develop spontaneously with increasing hole doping.
The crossover tracks the changing relative strengths of the intralayer and interlayer spin correlations.
Such competing correlations are directly relevant to \ce{La3Ni2O7}: substantial antiferromagnetic interlayer exchange has been inferred from neutron-scattering and RIXS measurements~\cite{327spinfulcChen2026,327SDWRIXSChen2024}, and stripe magnetic order together with stripe-related excitations has recently been observed~\cite{327spinfulcChen2026}.
How these magnetic correlations connect to the superconducting pairing remains an open question, given the experimental challenges of probing magnetism in the superconducting pressure regime.
Although a complete understanding of superconductivity in \ce{La3Ni2O7} will require multiorbital physics and structural effects beyond our minimal model, our results establish the structure of spin fluctuations as a key link between spin stripes and pairing competition, providing a minimal-model baseline for interpreting the magnetic excitations of bilayer nickelates.

\paragraph{Acknowledgments---}\hspace{-1em}
This work was supported by National Natural Science Foundation of China (Grants No. W2511008, 12088101, 12474218, 12674178), MOST (Grant Nos. 2022YFA1402400, 2022YFA1402700), and Science Challenge Project (Grant No. TZ2025013).

\paragraph{Data availability---}\hspace{-1em}
The data are not publicly available. The data are available from the authors upon reasonable request.

\bibliography{reference}

\onecolumngrid
\begin{center}
\textbf{End Matter}
\end{center}

\begin{figure}[H]
\centering
\includegraphics[width=\textwidth]{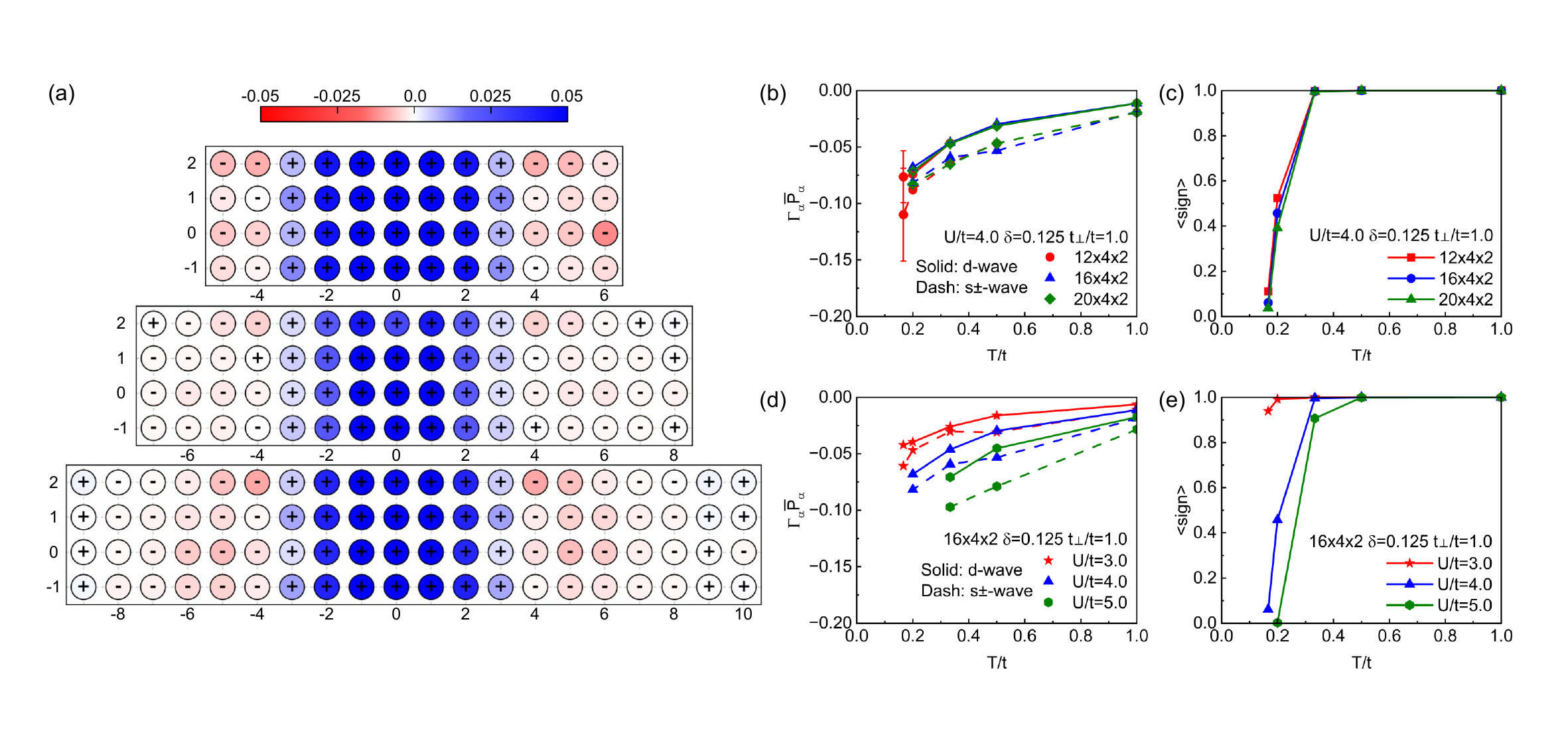}
\caption{
(a) Staggered real-space spin susceptibility $\chi^{\mathrm{spin}}_{11}(\mathbf r)$ for $12\times4\times2$, $16\times4\times2$, and $20\times4\times2$ clusters (top to bottom) at $U/t=4$ and $T/t=0.2$. 
The two layers give nearly identical results, so only layer 1 is shown.
Blue (red) circles marked with $+$ ($-$) denote positive (negative) values, and the three rows share the same color scale. 
(b) Temperature dependence of the effective pairing interaction $\Gamma_\alpha\overline P_\alpha$ of the $d$-wave (solid lines) and $s_\pm$-wave (dashed lines) and (c) the average fermion sign, for the three cluster sizes in (a). 
(d) Temperature dependence of the effective pairing interaction and (e) the average fermion sign, for $U/t=3$, $4$, and $5$ on the $16\times4\times2$ cluster.
All panels use $\delta=0.125$, $t_\perp/t=1.0$, and $t'/t=-0.25$. Error bars are omitted when they are smaller than the symbols.
}
\label{fig5}
\end{figure}
\twocolumngrid

We first compare different cluster lengths at $\delta=0.125$ to examine the size dependence of the stripe and pairing features discussed in the main text. 
As shown in Fig.~\ref{fig5}(a), all three clusters exhibit similar antiphase structures with comparable domain-wall spacings. 
Figure~\ref{fig5}(b) further shows that the effective pairing interactions of the $d$-wave and $s_\pm$-wave pairing have only modest size dependence. 
Meanwhile, the average fermion sign decreases rapidly with decreasing temperature, and larger clusters further aggravate the sign problem [Fig.~\ref{fig5}(c)]. 
These comparisons justify our choice of the $16\times4\times2$ cluster, which accommodates the stripe structure while remaining computationally feasible.

We next examine the effects of interaction strength and temperature in Fig.~\ref{fig5}(d) and \ref{fig5}(e), obtained on the $16\times4\times2$ cluster. 
For both $d$-wave and $s_\pm$-wave pairing, the effective pairing interactions generally become more attractive with decreasing temperature, and larger $U$ strengthens the attraction at the same temperature [Fig.~\ref{fig5}(d)]. 
These trends, however, are accompanied by a more severe sign problem [Fig.~\ref{fig5}(e)]: the average fermion sign approaches zero for $U/t=5$ at $T/t=0.2$, precluding reliable sampling in that regime. 
These comparisons motivate our choice of $U/t=4$ and $T/t=0.2$ for the calculations presented in the main text, and chart the accessible parameter regime for future simulations at stronger interactions and lower temperatures.

\end{document}